\documentclass[oupdraft]{bio}

\usepackage{amsmath}
\usepackage{amssymb}
\usepackage{url}
\usepackage{natbib}
\usepackage[figuresright]{rotating}
\usepackage[english]{babel}
\usepackage{multirow}
\usepackage{booktabs}
\usepackage{hhline}
\usepackage{bm}
\usepackage{enumerate}
\usepackage{verbatim}
\usepackage{color}

\newcommand{\bfY}{\bm{Y}}
\newcommand{\bfD}{\bm{D}}
\newcommand{\bfmu}{\bm{\mu}}
\newcommand{\bfeta}{\bm{\eta}}

\newcommand{\bftheta}{\bm{\theta}}

\newcommand{\bfone}{\bm{1}}
\newcommand{\bfI}{\bm{I}}
\newcommand{\bfV}{\bm{V}}

\newcommand{\bfR}{\bm{R}}

\newcommand{\bfB}{\bm{B}}
\newcommand{\bfE}{\bm{E}}
\newcommand{\bfM}{\bm{M}}
\newcommand{\bfS}{\bm{S}}
\newcommand{\bfH}{\bm{H}}
\newcommand{\bfQ}{\bm{Q}}
\newcommand{\bfP}{\bm{P}}
\newcommand{\bfC}{\bm{C}}
\newcommand{\bfPsi}{\bm{\Psi}}

\newcommand{\bfOmega}{\bm{\Omega}}
\newcommand{\bfLambda}{\bm{\Lambda}}
\newcommand{\bfalpha}{\bm{\alpha}}

\newcommand{\bfzero}{\bm{0}}
\newcommand{\bfvartheta}{\bm{\vartheta}}

\newcommand\given[1][]{\:#1\vert\:}

\begin{document}

\title{Marginal Modeling of Cluster-Period Means and Intraclass Correlations in Stepped Wedge Designs with Binary Outcomes}

\author{FAN LI$^{1,2,\ast}$, HENGSHI YU$^3$, PAUL J. RATHOUZ$^4$, ELIZABETH L. TURNER$^5$,\\
JOHN S. PREISSER$^6$\\[4pt]
\textit{$^{1}$Department of Biostatistics, Yale School of Public Health, New Haven, CT, U.S.A.\\
$^{2}$Center for Methods in Implementation and Prevention Science, Yale School of Public Health, New Haven, CT, U.S.A.\\
$^{3}$Department of Biostatistics, University of Michigan, Ann Arbor, MI, U.S.A.\\
$^{4}$Department of Population Health, The University of Texas at Austin, Austin, TX, U.S.A.\\
$^{5}$Department of Biostatistics and Bioinformatics, Duke University, Durham, NC, U.S.A.\\
$^{6}$Department of Biostatistics, University of North Carolina, Chapel Hill, NC, U.S.A.\\}
{*fan.f.li@yale.edu}}

\markboth%
{F. Li and others}
{Marginal Modeling of SW Trials with Binary Outcomes}

\maketitle

\footnotetext{To whom correspondence should be addressed.}

\begin{abstract}
{Stepped wedge cluster randomized trials (SW-CRTs) with binary outcomes are increasingly used in prevention and implementation studies. Marginal models represent a flexible tool for analyzing SW-CRTs with population-averaged interpretations, but the joint estimation of the mean and intraclass correlation coefficients (ICCs) can be computationally intensive due to large cluster-period sizes. Motivated by the need for marginal inference in SW-CRTs, we propose a simple and efficient estimating equations approach to analyze cluster-period means. We show that the quasi-score for the marginal mean defined from individual-level observations can be reformulated as the quasi-score for the same marginal mean defined from the cluster-period means. An additional mapping of the individual-level ICCs into correlations for the cluster-period means further provides a rigorous justification for the cluster-period approach. The proposed approach addresses a long-recognized computational burden associated with estimating equations defined based on individual-level observations, and enables fast point and interval estimation of the intervention effect and correlations. We further propose matrix-adjusted estimating equations to improve the finite-sample inference for ICCs. By providing a valid approach to estimate ICCs within the class of generalized linear models for correlated binary outcomes, this article operationalizes key recommendations from the CONSORT extension to SW-CRTs, including the reporting of ICCs.}
{Cluster randomized trials; Generalized estimating equations; Matrix-adjusted estimating equations (MAEE); Intraclass correlation coefficient; Statistical efficiency; Finite-sample correction.}
\end{abstract}
\section{Introduction}
\subsection{Overview and Objectives}\label{sec:overview}
Cluster randomized trials (CRTs) are pragmatic clinical trials that test interventions applied to groups or clusters \citep{Hayes2009}. Methodology for designing, conducting and analyzing CRTs has been rigorously developed over decades \citep{Turner2017a,Turner2017b}. A principal, but not the sole reason why CRTs are considered is that the intervention has one or more components defined at the cluster level. Increasingly, CRTs employ stepped wedge (SW) designs, which are one-way crossover designs where all clusters start out in the control condition and switch to the intervention at randomly assigned time points \citep{Hussey2007}. Logistical considerations such as the need to deliver the intervention in stages and the desire to eventually implement the intervention in all clusters are key factors involved in the decision to adopt a SW-CRT. Given the increasing popularity of these designs, the development of statistical methods and computational tools for valid analysis is critically important.

For the past decade, the design and analysis of SW-CRTs have mostly been based on linear mixed models \citep{LiReview2020}. Particularly, a major direction of research has been to study these methods under different random effects structures whose choice induces a marginal covariance structure \citep{Hooper2016,Kasza2017}. While not as frequently studied in the SW-CRT literature, generalized linear mixed models (GLMM) are a broad class of cluster-specific models to analyze clustered binary outcomes. However, their application carries a couple of caveats. First, with few exceptions (e.g., identity link), the interpretation of the intervention effect changes according to different specifications of the latent random-effects structure. Second, while GLMMs are flexible insofar as accounting for the dependence of observations within clusters via random effects, they may not adequately describe the pattern and magnitude of intraclass correlation structures on the natural measurement scale of the outcomes. This is because exact expressions for the marginal mean and correlation are generally lacking for GLMMs with a non-identity link function \citep{Zeger1988}. Perhaps for this reason, while GLMMs are used in the analysis of SW-CRTs with binary outcomes, they are seldom used as the basis for planning SW-CRTs, with an exception in \citet{Zhou2020} who developed a numerical approach for power calculation using the random-intercept linear probability model.

Motivated by the Washington State Expedited Partner Therapy (EPT) trial \citep{Golden2015}, we consider marginal model based analyses of SW-CRTs with binary outcomes and flexible choice of link functions. Marginal models separately specify the mean and the intraclass correlation structures, with the interpretation of the marginal mean parameters remaining the same regardless of correlation specification \citep{Liang1986,Zeger1988}. Further, the marginal modeling approach may be more robust because misspecification of the correlation structure does not affect the consistency of the regression parameter estimator in the marginal mean model. Finally, the marginal modeling framework permits direct estimation of intraclass correlation coefficients (ICCs) and assessing their uncertainty on the natural measurement scale of the outcomes. Such information is particularly useful as input parameters for sample size determination in cluster trials, generally \citep{Preisser2003}, and in SW-CRTs, specifically \citep{LiTurnerPreisser2018}. Accurate reporting of the intraclass correlation structures has been long advocated in parallel CRTs \citep{Preisser2007} and aligns with item 17a of the recent CONSORT extension to SW-CRTs, which recommends reporting of various intraclass correlation estimates to facilitate the planning of future trials \citep{Hemming2018}.

\subsection{Motivating Study: The Washington State EPT Trial}\label{sec:EPT}
The Washington State EPT trial is a SW-CRT that evaluates the population effect of an expedited patient-delivered partner notification strategy versus the standard partner notification for the treatment of Chlamydia and Gonorrhea infection \citep{Golden2015}. The intervention includes the promotion of patient-delivered partner therapy through commercial pharmacies and targeted provision of public health partner services, and was designed to increase treatment adoption for sex partners of individual heterosexual patients. The randomization is carried out at the level of local health jurisdiction (LHJ), namely the administrative unit corresponding to a single county. Each LHJ is a cluster, and a total of 23 LHJs were randomized from 2007 to 2010 over four waves until the intervention had been disseminated in all LHJs. Cross-sectional surveys were conducted based on sentinel women aged 14 to 25 years in each LHJ at baseline and in between waves to measure the prevalence of Chlamydia and Gonorrhea. Due to the cross-sectional design, different women are included in different periods.

\begin{figure}[!p]
\centering\includegraphics[scale=0.6]{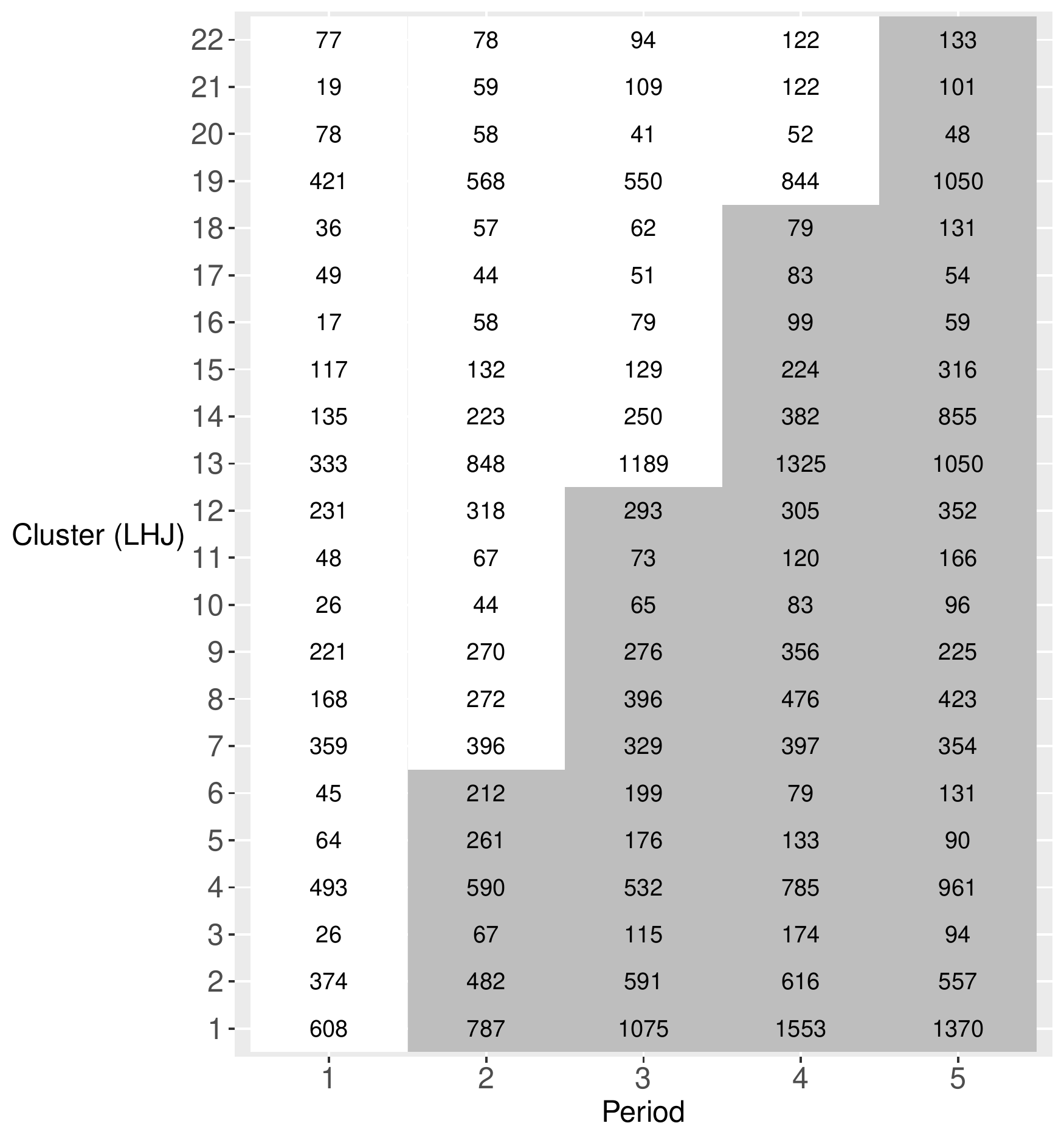}
\caption{Cluster-by-period diagram of the Washington State EPT Trial. Each white cell represents a control cluster-period and each gray cell indicates an intervention cluster-period. The corresponding cluster-period sizes are indicated in each cell.}
\label{fig:Fig1}
\end{figure}

Following \citet{Golden2015}, we restrict the analysis to the $22$ LHJs that provide individual-level data on the Chlamydia outcome. Define $Y_{ijk}$ as the binary Chlamydia infection status for sentinel woman $k=1,\ldots,n_{ij}$ surveyed during period $j=1,\ldots,J$ in LHJ $i=1,\ldots,I$; the value of $Y_{ijk}$ equals $1$ if the sentinel woman reports Chlamydia and $0$ otherwise. \citet{LiTurnerPreisser2018} and \citet{Li2020} specified the following marginal mean model for SW-CRTs
\begin{equation}\label{eq:mm}
g(\mu_{ijk})=\beta_j+X_{ij}\delta,
\end{equation}
where $\mu_{ijk}$ is the mean of $Y_{ijk}$, $g$ is the link function, $\beta_j$ is the $j$th period effect, $X_{ij}$ is the intervention indicator, and $\delta$ is the time-adjusted average intervention effect on the link function scale. Because it is also of interest to report the within-period and between-period correlations, \citet{LiTurnerPreisser2018} proposed a paired estimating equations approach to simultaneously estimate the intervention effect and correlation parameters. However, in many cross-sectional SW-CRTs, the cluster-period sizes $n_{ij}$'s are large and highly variable. In the EPT trial design, for example, the cluster-by-period diagram in Figure \ref{fig:Fig1} shows that the cluster size $n_{i+}=\sum_{j=1}^J n_{ij}$ ranges from $277$ to $5393$. As the correlation estimating equations can involve as many as ${5393\choose 2}\approx 14.5$ million residual cross-product terms in one cluster, the estimation of correlations via individual-level analysis quickly becomes computationally burdensome or infeasible.

With these considerations in mind, this article develops a computationally convenient marginal modeling approach to estimate the intervention effect, the ICCs and their respective sampling variances in SW-CRTs with large and variable cluster sizes. While binary outcomes are the focus of the methods development, application and evaluation of statistical properties of the proposed method, extensions to continuous and count outcomes are discussed in the Supplementary Material (SM). In what follows, we reformulate the estimating equations based on individual-level analysis to ones based on cluster-period means to alleviate the computational challenge without compromising the ability to estimate individual-level ICCs. We allow flexible link functions for the marginal mean and consider the estimation and inference for two multilevel correlation structures appropriate for cross-sectional SW designs.

\section{Modeling Cluster-Period means in Cross-Sectional Designs}\label{sec:GEE}

Because marginal mean model \eqref{eq:mm} is a function of both period and intervention, we consider collapsing the individual-level outcomes to cluster-period means $\overline{\bfY}_i=(\overline{Y}_{i1},\ldots,\overline{Y}_{iJ})^T=(Y_{i1+}/n_{i1},\ldots,Y_{iJ+}/n_{iJ})^T$, where $Y_{ij+}=\sum_{k=1}^{n_{ij}}Y_{ijk}$ is the cluster-period total, and $n_{ij}$ the cluster-period size. We assume that the cluster-period sizes are variable, which is almost always the case with cross-sectional SW-CRTs. Let the mean of $\overline{\bfY}_i$ be $\bfmu_i=(\mu_{i1},\ldots,\mu_{iJ})^T$, where for a binary outcome $\mu_{ij}=E(Y_{ij+})/n_{ij}$ is the prevalence in the $(i,j)$th cluster-period. Because the right hand side of marginal model \eqref{eq:mm} depends only on cluster and period, aggregating over cluster-periods implies the same marginal model, $g(\mu_{ij})=\beta_j+X_{ij}\delta$. Writing $\bftheta=(\beta_1,\ldots,\beta_J,\delta)^T$, the generalized estimating equations \citep[GEE;][]{Liang1986} for $\bftheta$ are
\begin{equation}\label{eq:gee}
\sum_{i=1}^I \bfD_{1i}^T\bfV_{1i}^{-1}(\overline{\bfY}_i-\bfmu_i)=0,
\end{equation}
where $\bfD_{1i}=\partial\bfmu_i/\partial\bftheta^T$ and $\bfV_{1i}=\text{cov}(\overline{\bfY}_i)$ is the working covariance matrix parameterized by the individual-level variances and pairwise correlations. This is the usual GEE applied to correlated binomial data, and our novel contribution is to enable the estimation of individual-level ICCs that parameterize the cluster-period mean covariance $\bfV_{1i}$.

Several prior efforts collapsed individual-level observations for analyzing SW-CRTs. \citet{Hussey2007} suggested a linear mixed model based on cluster-period means with a random intercept. For binary outcomes, their approach only estimates the treatment effect on the risk difference scale and does not estimate valid ICCs defined from the individual-level model under variable cluster sizes (see SM Appendix A for details). \citet{Thompson2018} proposed a permutation test based on cluster-period means. However, their approach assumed working independence and ignored the estimation of correlation structures. Our approach distinguishes from these two earlier efforts by allowing arbitrary link functions in the marginal mean model for binary outcomes as well as by enabling valid estimation and inference for ICC structures.

In cross-sectional designs, distinct sets of participants are included in each period, and require modeling both the within-period and between-period correlations for each pair of individual-level outcomes. We consider two multilevel correlation structures: the nested exchangeable and the exponential decay structures. The nested exchangeable correlation structure \citep{LiTurnerPreisser2018} differentiates between the within-period and between-period ICCs. Specifically, this structure assumes a constant correlation $\alpha_0$ between two individual outcomes from the same cluster within the same period, and a constant correlation $\alpha_1$ between two individual outcomes from the same cluster across two periods. Equating $\alpha_1$ with $\alpha_0$ leads to the simple exchangeable structure as in standard GEE analyses \citep{Hussey2007}. The exponential decay correlation structure was recently introduced in the context of linear mixed models \citep{Kasza2017,LiReview2020}. While this structure assumes a constant correlation $\alpha_0$ between two individual outcomes from the same cluster within the same period, it allows the between-period correlation to decay at an exponential rate. Mathematically, the correlation between two outcomes measured in the $j$th and $l$th periods ($1\leq j,l\leq J$) is $\alpha_0\rho^{|j-l|}$ ($0\leq \rho\leq 1$). In the absence of decay ($\rho=1$), the exponential decay structure also reduces to the simple exchangeable structure. Example matrix forms of these correlation structures are provided in STable 1 in the SM. In the following two sections, we develop estimation and inference strategies under each of these correlation structures.

\subsection{Nested Exchangeable Correlation Structure}\label{sec:nex}
The individual-level correlation structure informs the specification of covariances for cluster-period means. Under the nested exchangeable correlation structure, the diagonal element of $\bfV_{1i}$ is
\begin{equation}\label{eq:sii}
\sigma_{ijj}=\text{var}(\overline{Y}_{ij+})=\frac{\nu_{ij}}{n_{ij}}
\left\{1+(n_{ij}-1)\alpha_0\right\},
\end{equation}
where $\nu_{ij}=\mu_{ij}(1-\mu_{ij})$ is the binomial variance. The design effect, $1+(n_{ij}-1)\alpha_0$, is the classic variance inflation factor for over-dispersed binomial outcomes. The off-diagonal element of $\bfV_{1i}$ is
\begin{equation}
\sigma_{ijl}=\text{cov}(\overline{Y}_{ij+},\overline{Y}_{il+})
=\sqrt{\nu_{ij}\nu_{il}}\alpha_1.
\end{equation}
When all $n_{ij}\rightarrow\infty$, $\text{var}(\overline{Y}_{ij+})\rightarrow \nu_{ij}\alpha_0$ and the pairwise cluster-period mean correlation $\text{corr}(\overline{Y}_{ij+},\overline{Y}_{il+})\rightarrow {\alpha_1}/{\alpha_0}$, which is identical to the cluster autocorrelation defined in \citet{Hooper2016} and \citet{LiReview2020} based on linear mixed models. This also suggests that the cluster-period means are approximately exchangeable when all $n_{ij}$'s are large, but such an approximation may be crude in cases such as the motivating study where the $n_{ij}$'s vary from $19$ to $1553$.

Define $\bfalpha=(\alpha_0,\alpha_1)^T$, we specify the covariance estimating equations \citep{Zhao1990} to estimate $\bfalpha$
\begin{equation}\label{eq:gamma}
\sum_{i=1}^I \bfD_{2i}^T\bfV_{2i}^{-1}(\bfS_i-\bfeta_i)=0,
\end{equation}
where $\bfeta_i=(\sigma_{i11},\sigma_{i12},\ldots,\sigma_{i22},\ldots)^T$, $\bfS_i=(s_{i11},s_{i12},\ldots,s_{i22},\ldots)^T$, $s_{ijl}=(\overline{Y}_{ij+}-\hat{\mu}_{ij})(\overline{Y}_{il+}-\hat{\mu}_{il})$ is the residual cross-product, $\bfD_{2i}=\partial \bfeta_i/\partial\bfalpha^T$ and $\bfV_{2i}$ is the working variance for $\bfS_i$. Parametric specification of working covariances $\bfV_{2i}$ requires the joint distributions of within-cluster triplets and quartets, which are not provided from the specification of marginal mean and covariances. Henceforth, a practical strategy is to set $\bfV_{i2}$ as identity matrix \citep{Sharples1992}. In this case, the following closed-form updates are implied from \eqref{eq:gamma}
\begin{align}\label{eq:solution}
\hat{\alpha}_0=\frac{\sum_{i=1}^I\sum_{j=1}^{J}\left(\frac{n_{ij}-1}{n_{ij}}\right)\hat{\nu}_{ij}
\left(s_{ijj}-\frac{\hat{\nu}_{ij}}{n_{ij}}\right)}
{\sum_{i=1}^I\sum_{j=1}^{J}\left(\frac{n_{ij}-1}{n_{ij}}\right)^2\hat{\nu}_{ij}^2},
~~~~\hat{\alpha}_1=\frac{\sum_{i=1}^I\sum_{j\neq l}s_{ijl}\sqrt{\hat{\nu}_{ij}\hat{\nu}_{il}}}{\sum_{i=1}^I\sum_{j\neq l} \hat{\nu}_{ij}\hat{\nu}_{il}}.
\end{align}

Noticeably, even though the cluster-period sizes $n_{ij}$ could be large and pose a computational challenge for individual-level paired estimating equations, cluster-period aggregation reduces the effective cluster sizes to $J$, the number of periods, which rarely exceeds $10$ \citep{Grayling2017b}. On the other hand, because SW-CRTs often involve a limited number of clusters (fewer than $30$), the residual vector $\overline{\bfY}_i-\hat{\bfmu}_i$ could be biased towards zero due to overfitting, leading to finite-sample bias in the estimation of correlation parameters. Here we extend the multiplicative adjustment of \citet{Preisser2008} to the covariance estimating equations \eqref{eq:gamma} by the following argument. Because $E[(\overline{\bfY}_{i}-\hat{\bfmu}_i)(\overline{\bfY}_{i}-\hat{\bfmu}_i)^T]\approx (\bfI-\bfH_{1i})\text{cov}(\overline{\bfY}_i)$, where $\bfH_{1i}=\bfD_{1i}(\sum_{i=1}^I\bfD_{1i}\bfV_{1i}^{-1}\bfD_{1i})^{-1}\bfD_{1i}^T\bfV_{1i}$ is the cluster leverage, a bias-adjusted, and hence more accurate estimator for the covariance of $\overline{\bfY}_i$ is obtained as
\begin{equation}\label{eq:correction}
\widetilde{\text{cov}}(\overline{\bfY}_i)=(\bfI-\bfH_{1i})^{-1}(\overline{\bfY}_{i}-\hat{\bfmu}_i)(\overline{\bfY}_{i}-\hat{\bfmu}_i)^T,
\end{equation}
where $\bfH_{1i}$ is evaluated at $\hat{\bftheta}$. Improved estimation of correlation parameters may then be achieved by replacing $\bfS_i$ in \eqref{eq:gamma} with $\widetilde{\bfS}_i=(\tilde{s}_{i11},\tilde{s}_{i12},\ldots,\tilde{s}_{i22},\ldots)$, where $\tilde{s}_{ijl}$ is the $(j,l)$th element of the bias-adjusted covariance $\widetilde{\text{cov}}(\overline{\bfY}_i)$. We will similarly define the cluster leverage for the covariance estimating equations as $\bfH_{2i}=\bfD_{2i}(\sum_{i=1}^I\bfD_{2i}\bfV_{2i}^{-1}\bfD_{2i})^{-1}\bfD_{2i}^T\bfV_{2i}$, which is evaluated at $\hat{\bftheta}$ and $\hat{\bfalpha}$.

When the number of clusters $I$ is large, the joint distribution of $I^{1/2}(\hat{\bftheta}-\bftheta)$, $I^{1/2}(\hat{\bfalpha}-\bfalpha)$ is Gaussian with mean zero and covariances estimated by
\begin{equation}\label{eq:cov}
I\times \left(
\begin{array}{cc}
\bfOmega & \bfzero \\
\bfQ & \bfP \\
\end{array}
\right)\left(
\begin{array}{cc}
\bfLambda_{11} & \bfLambda_{12} \\
\bfLambda_{12}^T & \bfLambda_{22} \\
\end{array}
\right)\left(
\begin{array}{cc}
\bfOmega & \bfQ^T \\
\bfzero & \bfP \\
\end{array}
\right),
\end{equation}
where $\bfOmega=\left\{\sum_{i=1}^I\bfD^T_{1i}\bfV_{1i}^{-1}\bfD_{1i}\right\}^{-1}$, $\bfP=\left\{\sum_{i=1}^I\bfD^T_{2i}\bfV_{2i}^{-1}\bfD_{2i}\right\}^{-1}$, $\bfQ=\bfP\left\{\sum_{i=1}^I \bfD^T_{2i}\bfV_{2i}^{-1}\frac{\partial \bfS_i}{\partial\bftheta^T}\right\}\bfOmega$,
\begin{align*}
\bfLambda_{11}&=\sum_{i=1}^I \bfC_{1i}\bfD_{1i}^T\bfV_{1i}^{-1}\bfB_{1i}(\overline{\bfY}_i-\hat{\bfmu}_i)
(\overline{\bfY}_i-\hat{\bfmu}_i)^T\bfB_{1i}^T\bfV_{1i}^{-1}\bfD_{1i}\bfC_{1i}^T\\
\bfLambda_{12}&=\sum_{i=1}^I \bfC_{1i}\bfD_{1i}^T\bfV_{1i}^{-1}\bfB_{1i}(\overline{\bfY}_i-\hat{\bfmu}_i)
(\tilde{\bfS}_i-\hat{\bfeta}_i)^T\bfB_{2i}^T\bfV_{2i}^{-1}\bfD_{2i}\bfC_{2i}^T\\
\bfLambda_{22}&=\sum_{i=1}^I \bfC_{2i}\bfD_{2i}^T\bfV_{2i}^{-1}\bfB_{2i}(\tilde{\bfS}_i-\hat{\bfeta}_i)
(\tilde{\bfS}_i-\hat{\bfeta}_i)^T\bfB_{2i}^T\bfV_{2i}^{-1}\bfD_{2i}\bfC_{2i}^T,
\end{align*}
where we will discuss the choice of $\{\bfC_{1i},\bfC_{2i}\}$ and $\{\bfB_{1i},\bfB_{2i}\}$ in the following. If we set $\bfC_{1i}=\bfI_{\text{dim}(\bftheta)}$, $\bfC_{2i}=\bfI_{\text{dim}(\bfalpha)}$, $\bfB_{1i}=\bfI_{\text{dim}(\overline{\bfY}_i)}$, $\bfB_{2i}=\bfI_{\text{dim}(\tilde{\bfS}_i)}$, equation \eqref{eq:cov} becomes the robust sandwich variance in the spirit of \citet{Zhao1990}, or BC0. Because the number of clusters included in SW-CRTs are frequently less than $30$, the following finite-sample bias corrections could provide improved inference for $\bftheta$ and $\bfalpha$. Specifically, setting $\bfC_{1i}$, $\bfC_{2i}$ as identity but $\bfB_{1i}=(\bfI_{\text{dim}(\overline{\bfY}_i)}-\bfH_{1i})^{-1/2}$, $\bfB_{2i}=(\bfI_{\text{dim}(\tilde{\bfS}_i)}-\bfH_{2i})^{-1/2}$ results in the bias-corrected covariance that extends \citet{Kauermann2001}, or BC1. Setting $\bfC_{1i}$, $\bfC_{2i}$ as identity but $\bfB_{1i}=(\bfI_{\text{dim}(\overline{\bfY}_i)}-\bfH_{1i})^{-1}$, $\bfB_{2i}=(\bfI_{\text{dim}(\tilde{\bfS}_i)}-\bfH_{2i})^{-1}$ results in the bias-corrected covariance that extends \citet{Mancl2001}, or BC2. Finally, setting $\bfB_{1i}$, $\bfB_{2i}$ as identity but $\bfC_{1i}=\text{diag}\{(1-\min\{\zeta_1,[\bfD_{1i}^T\bfV_{1i}^{-1}\bfD_{1i}]_{jj}\bfOmega\})^{-1/2}\}$, $\bfC_{2i}=\text{diag}\{(1-\min\{\zeta_2,[\bfD_{2i}^T\bfV_{2i}^{-1}\bfD_{2i}]_{jj}\bfP\})^{-1/2}\}$
extends \citet{Fay2001}, or BC3. Usually we set $\zeta_1=\zeta_2=0.75$ to ensure that multiplicative bias correction is no larger than 2 fold. When $I$ is smaller than 30, each of these bias-corrections could inflate the variance relative to BC0 and potentially improve the finite-sample behaviour of the sandwich variance.

\subsection{Exponential Decay Correlation Structure}\label{sec:ed}
Under the exponential decay correlation structure, the covariances for cluster-period means $\bfV_{1i}$ include diagonal element $\sigma_{ijj}$ defined in equation \eqref{eq:sii}, and off-diagonal element becomes $\sigma_{ijl}=\text{cov}(\overline{Y}_{ij+},\overline{Y}_{il+})
=\sqrt{\nu_{ij}\nu_{il}}\alpha_0\rho^{|j-l|}$. When all $n_{ij}\rightarrow\infty$, the pairwise cluster-period mean correlation $\text{corr}(\overline{Y}_{ij+},\overline{Y}_{il+})\rightarrow\rho^{|j-l|}$, which corresponds to a first-order auto-regressive structure. Again, such an approximation may not be accurate in the Washington State EPT trial because the cluster-period sizes could occasionally be small and quite variable.

Unlike the expression obtained under the nested exchangeable correlation structure, $\sigma_{ijl}$ obtained under the exponential decay structure is nonlinear in the decay parameter $\rho$. Based on estimating equations \eqref{eq:gamma} and the bias-adjusted covariance $\widetilde{\bfS}_i$, we can show that each update of $(\alpha_0,\rho)$ joint solves the following system of equations
\begin{align}
&\hat{\alpha}_0=\frac{\sum_{i=1}^I\sum_{j=1}^J \left(\frac{n_{ij}-1}{n_{ij}}\right)\left(\tilde{s}_{ijj}\hat{\nu}_{ij}-\frac{\hat{\nu}^2_{ij}}{n_{ij}}\right)
+\sum_{i=1}^I\sum_{j\neq l}\tilde{s}_{ijl}\sqrt{\hat{\nu}_{ij}\hat{\nu}_{il}}\hat{\rho}^{|j-l|}}
{\sum_{i=1}^I\sum_{j=1}^{J}\left(\frac{n_{ij}-1}{n_{ij}}\right)^2\hat{\nu}_{ij}^2
+\sum_{i=1}^I\sum_{j\neq l}\hat{\nu}_{ij}\hat{\nu}_{il}\hat{\rho}^{2|j-l|}}
\label{eq:eea0}\\
&\sum_{i=1}^I\sum_{j\neq l}|j-l|\tilde{s}_{ijl}\sqrt{\hat{\nu}_{ij}\hat{\nu}_{il}}\hat{\rho}^{|j-l|-1}
-\sum_{i=1}^I\sum_{j\neq l}|j-l|\hat{\nu}_{ij}\hat{\nu}_{il}\hat{\alpha}_0\hat{\rho}^{2|j-l|-1}=0.\label{eq:eerho}
\end{align}
In particular, we observe that the second equation is a polynomial function of $\rho$ to the order of $2|J-1|-1$, and so one can use root-finding algorithms to search for the zero-value within the unit interval. Given each update of the marginal mean parameters, an update of the exponential decay correlation structure can be obtained by iterating between \eqref{eq:eea0} and \eqref{eq:eerho}. The variance estimators for both $\bftheta$ and $\bfalpha$ with finite-sample corrections can be obtained by following the approach in Section \ref{sec:nex}. Extensions of the proposed cluster-period marginal modeling approach for continuous and count outcomes are presented in the SM Appendix B.

\section{Considerations on Asymptotic Efficiency}\label{sec:eff}

We assess the asymptotic efficiency in estimating the intervention effect $\delta$ based on estimating equations defined for cluster-period means. In the same context, \citet{LiTurnerPreisser2018} provided a paired estimating equations for individual-level outcomes, which in principle serves as the efficiency gold-standard. However, the computational burdens of that approach in analyzing the motivating trial are twofold: those associated with repeatedly inverting a large correlation matrix for marginal mean estimation and those associated with enumerating all pairwise residual cross-products for correlation estimation. These computational disadvantages prohibit the application of individual-level GEE to analyze SW-CRTs with large cluster-period sizes, especially when the correlation model includes more than one parameter. In contrast, the cluster-period GEE converges in seconds because the induced correlation matrix is of dimension $J\times J$ and only $J\choose 2$ pairwise residual products need to be enumerated in each cluster. It is then of interest to study whether the cluster-period GEE compromises efficiency in estimating the intervention effect $\delta$. To proceed, we observe that both the nested exchangeable and exponential decay correlation structures are special cases of the block Toeplitz structure defined in SM Appendix C. In SM Appendix D, we show that, as long as the working correlation model for individual-level data is block Toeplitz, the marginal mean estimating equations for cluster-period means are equivalent to those for individual-level outcomes. Specifically, we define the individual-level estimating equations for $\bftheta$ as $\sum_{i=1}^I \bfE_{1i}^T\bfM_{1i}^{-1}(\bfY_i-\bfvartheta_i)=\bfzero$, where $\bfY_i=(Y_{i11},Y_{i12},\ldots,Y_{i21},\ldots)^T$, $\bfvartheta_i=(\mu_{i1}\bfone_{n_{i1}}^T,\ldots,\mu_{iJ}\bfone_{n_{iJ}}^T)^T$, $\bfM_{1i}=\text{cov}(\bfY_i)$, and $\bfE_{1i}=\partial\bfvartheta_i/\partial\bftheta^T$, and the following result holds.
\begin{theorem}\label{thm:same}
Assuming marginal mean model is \eqref{eq:mm} and the individual-level working correlation is block Toeplitz, the quasi-score defined from the cluster-period means and the induced cluster-period mean covariance structure is identical to the quasi-score defined from the individual-level outcomes. That is, $\bfD_{1i}^T\bfV_{1i}^{-1}(\overline{\bfY}_i-\bfmu)=\bfE_{1i}^T\bfM_{1i}^{-1}(\bfY_i-\bfvartheta_i)
$ for each cluster $i$. Similarly, $\bfD_{1i}^T\bfV_{1i}^{-1}\bfD_{1i}=\bfE_{1i}^T\bfM_{1i}^{-1}\bfE_{1i}$.
\end{theorem}



To establish the above general result under variable cluster-period sizes, a mathematical induction argument (SM Appendix D) is necessary because an analytical inverse cannot be easily obtained for the block Toeplitz matrix. Theorem \ref{thm:same} indicates that there is no loss of asymptotic efficiency for estimating the intervention effect that results from cluster-period aggregation, as long as the induced cluster-period mean correlation matrix is properly specified. Particularly, assuming equal cluster-period sizes and a linear mixed model with Gaussian outcomes, \citet{Grantham2019} suggested that the linear mixed model based on the cluster-period summary results in no loss of information for estimating the treatment effect in SW-CRTs as long as the within-period observations are exchangeable. Theorem \ref{thm:same} generalizes their finding to GEE with arbitrary specification of link and variance functions and further relaxes their equal cluster-period size assumption.

Theorem \ref{thm:same} also provides a convenient device to numerically evaluate the asymptotic relative efficiency (ARE) between accurately modeling the cluster-period mean correlations versus using a working independence structure. To further support the application of the proposed approach versus using working independence in analyzing the motivating study, we calculate the ARE in estimating $\delta$ between these two approaches. To do so, we assume a SW-CRT with $22$ clusters and $5$ periods, where the randomization follows the cluster-by-period diagram in Figure \ref{fig:Fig1}. We set the true marginal mean model as equation \eqref{eq:mm} with a logit link. The period effect $\beta_j$'s are specified so that the outcome prevalence decreases from $25\%$ to $20\%$ in the absence of intervention, and the intervention effect corresponds to an odds ratio of $e^\delta=0.75$. To account for variable cluster-period sizes, we resample the cluster-period sizes from the motivating study and obtain $1000$ bootstrap replicates. For the $k$th bootstrap replicate, we obtain
\begin{equation*}\label{eq:RE}
\tau_k=\frac{\left[(\sum_{i=1}^I \bfD_{1i}^T\bfPsi_{1i}^{-1}\bfD_{1i})^{-1}
 (\sum_{i=1}^I \bfD_{1i}^T\bfPsi_{1i}^{-1}\bfV_{1i}\bfPsi_{1i}^{-1}\bfD_{1i})
 (\sum_{i=1}^I \bfD_{1i}^T\bfPsi_{1i}^{-1}\bfD_{1i})^{-1}\right]_{(J+1,J+1)}}{\left[(\sum_{i=1}^I \bfD_{1i}^T\bfV_{1i}^{-1}\bfD_{1i})^{-1}\right]_{(J+1,J+1)}},
\end{equation*}
where $\bfPsi_{1i}$ is a $J\times J$ diagonal matrix with the $j$th element as $\nu_{ij}/n_{ij}$ (i.e. working independence covariance model), and all parameters evaluated at the truth. The ARE is then estimated as $\sum_{k=1}^{1000} \tau_k/1000$. Table \ref{tb:MRE} presents the ARE under different true correlation models. It is evident that the efficiency gain from properly modeling the correlations is maximum ($\text{ARE}\approx 44$ when $\alpha_0=\alpha_1=0.1$) when the within-period ICC and between-period ICC are identical, assuming the latter does not exceed the former. The ARE decreases when the within-period ICC decreases, and also when the between-period ICC deviates from the within-period ICC. In the scenario where $\alpha_0=0.02$ and $\alpha_1=0.001$, modeling the correlations is still $60\%$ more efficient than ignoring the correlation structure in estimating $\delta$. These observations highlight the importance of correlation specification in SW-CRTs when the cluster-period sizes are highly variable.

\section{Simulation Studies}\label{sec:sim}
We conduct two sets of simulation experiments to assess the finite-sample operating characteristics of the cluster-period GEE for analyzing correlated binary outcomes in cross-sectional SW-CRTs. In the first simulation experiment, we focus on a limited number of clusters $I\in\{12,24,36\}$ with treatment sequences randomized across $J=5$ periods. We assume all clusters receive the control condition during the first period $J=1$, and an equal number of clusters cross over to intervention at each wave. We use the \citet{Qaqish2003} method to generate correlated individual-level binary outcomes. The true marginal mean model is given by \eqref{eq:mm}, where $g$ is a logit link and the effect size $\delta=\log(0.5)$. We assume the baseline prevalence of the outcome is $35\%$ and a gently decreasing time trend with $\beta_{j}-\beta_{j+1}=0.1\times (0.5)^j$ for $j\geq 1$. Both the nested exchangeable and the exponential decay correlation structures are considered in simulating the data. When the true correlation is nested exchangeable, we consider $(\alpha_0,\alpha_1)\in\{(0.03,0.015), (0.1,0.05)\}$, representing small to moderate within-period and between-period correlations previously reported \citep{Martin2016b}. When the true correlation is exponential decay, we consider $(\alpha_0,\rho)\in\{(0.03,0.8),(0.1,0.5)\}$ in accordance with values assumed in previous simulations for cohort stepped wedge designs \citep{Li2020}. In this first simulation experiment, we consider relatively large but more variable cluster-period sizes, randomly drawn from DiscreteUniform$(50,150)$. Here, the maximum total number of observations in each cluster is allowed to be $750$, and the individual-level GEE described in \citet{LiTurnerPreisser2018} becomes computationally burdensome to fit due to (1) the enumeration of a maximum of ${750\choose 2}=280,875$ pairwise residual cross-products in each cluster and (2) numerical inversion of a large correlation matrix in each modified Fisher-scoring update. We therefore only consider analyzing the simulated data via the cluster-period GEE with the correct specification of the marginal mean and induced correlation structure. {We simulate $3000$ data replicates, and study the percent relative bias and coverage probability in estimating the marginal intervention effect and correlation parameters.} The comparisons are made between the uncorrected estimating equations (UEE), namely equation \eqref{eq:gamma}, and the matrix-adjusted estimating equations (MAEE), namely equation \eqref{eq:gamma} but now with the bias-adjusted cross-products $\widetilde{\bfS}_i$.

Table \ref{tb:bias} summarizes the percent relative bias results. Overall, the bias of the intervention effect remains insensitive to bias corrections of the correlation estimating equations, corroborating the findings of \citet{Lu2007} for individual-level GEE. However, when the true correlation structure is nested exchangeable, MAEE substantially reduces the negative bias of UEE in estimating $\alpha_0$ and $\alpha_1$. When the true correlation structure is exponential decay, MAEE similarly reduces the negative bias of UEE in estimating the within-period correlation $\alpha_0$, but comes at a cost of slightly inflating the negative bias in estimating the decay parameter $\rho$, especially when $I=12$. This is because $\alpha_0$ and $\rho$ enter the polynomial estimating equation \eqref{eq:eerho} in a multiplicative fashion, while the updates for $\alpha_0$ and $\alpha_1$ under the nested exchangeable structure are nearly orthogonal. As the number of clusters increase to $I=24$ or $36$, both MAEE and UEE have negligible bias in estimating $\rho$, but MAEE still has notably smaller bias in estimating $\alpha_0$.

STable 2 in the SM summarizes the coverage probability for $\delta$. The confidence intervals (CIs) are constructed based on the $t_{I-2}$ quantiles as this approach has been shown to provide robust small-sample behaviour in previous simulations with individual-level GEE \citep{Li2020,ford2020maintaining}. In addition to the model-based variance and the usual sandwich variance, we examine three bias-corrected variances introduced in Section \ref{sec:nex}. {Based on a binomial model with $3000$ replicates, we consider the empirical coverage between $94.2\%$ and $95.8\%$ as close to nominal. STable 2 indicates that the CIs for $\delta$ constructed with the model-based variance or any of the bias-corrected variances generally provide close to nominal coverage, while those based on BC0 frequently lead to under coverage. STable 3 and 4 summarize the coverage probability of the correlation parameters (interval constructed based on the same $t_{I-2}$ distribution). Given the limited number of clusters and variable cluster sizes, the coverage of correlation parameters is frequently below nominal. However, MAEE can substantially improve the coverage of the correlations parameters. Throughout, the CIs constructed based on BC2 provide the best coverage for correlations, a finding that echoes \citet{Perin2017} with alternating logistic regressions.}

\begin{figure}[!p]
\centering
\includegraphics[scale=0.6]{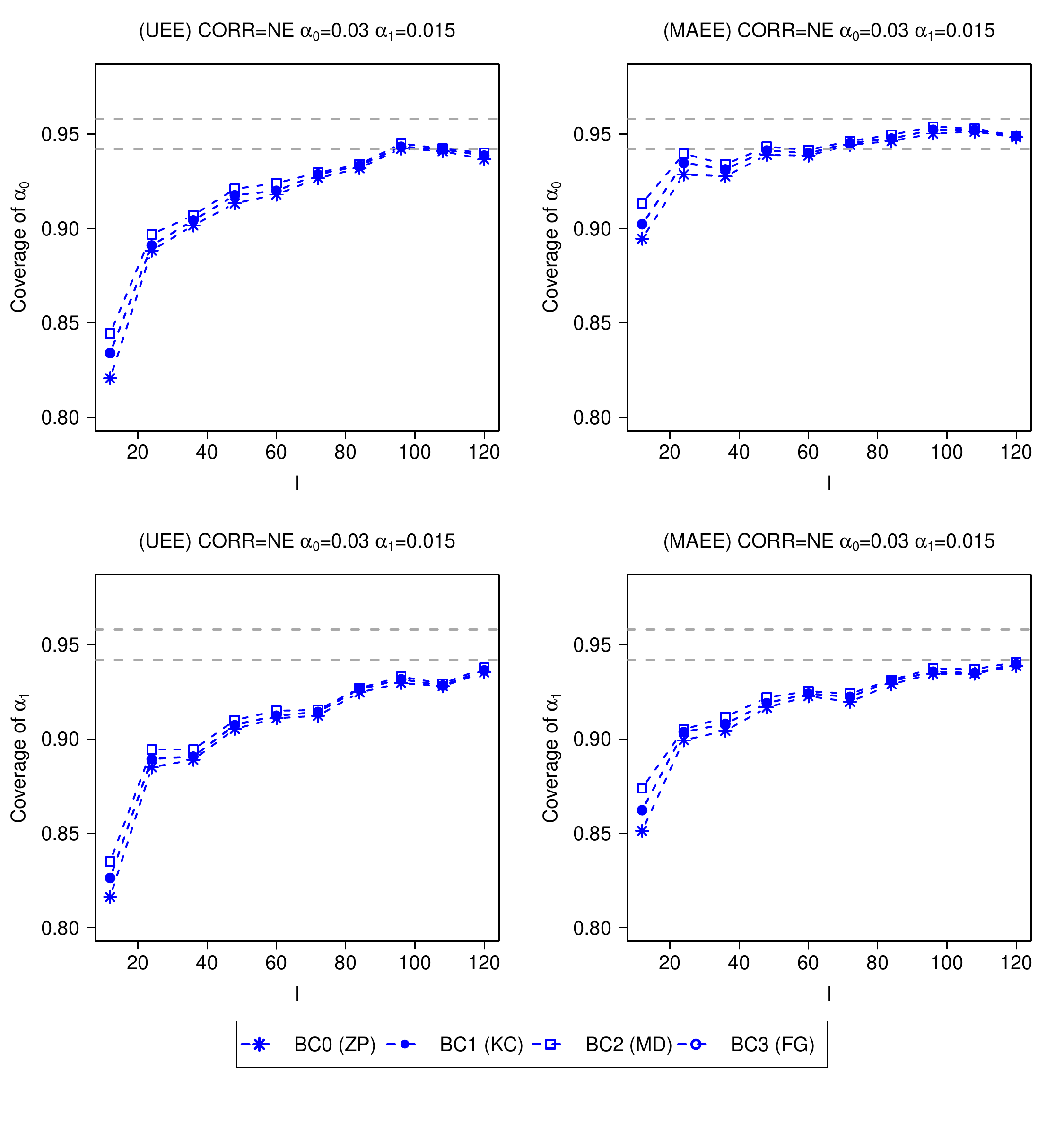}
\caption{Coverage of 95\% confidence intervals for correlation parameters based on the $t_{I-2}$ quantiles as a function of number of clusters $I$ under the nested exchangeable (NE) correlation structure when $\alpha_0=0.03$ and $\alpha_1=0.015$. The cluster-period sizes are randomly drawn from DiscreteUniform$(25,50)$. The acceptable bounds according to simulation error based on $3000$ replicates are shown with the dashed horizontal lines.}\label{fig:nex1cov}
\end{figure}

To further investigate the coverage probability of the correlation parameters with a larger number of clusters, we consider a second set of experiments, with the same simulation design except for smaller cluster-period sizes. Specifically, the cluster-period sizes are randomly drawn from DiscreteUniform$(25,50)$, and the number of clusters $I$ is varied from $12$ to $120$. The coverage results for the intervention effect parameter $\delta$ are largely consistent with those from the first simulation experiment, and are presented in SFigure 1 and 2 in the SM. {However, the results further indicate that the model-based variance and BC2 may lead to over coverage with $I=12$ and $I=24$, and that BC1 seems to have the most robust performance.} Next, Figure \ref{fig:nex1cov} presents the coverage for the nested exchangeable correlation structure when $\alpha_0=0.03$ and $\alpha_1=0.015$. Similar results for $\alpha_0=0.1$ and $\alpha_1=0.05$ are in SFigure 3. These figures indicate that MAEE coupled with BC2 leads to higher and closer to nominal coverage for $\alpha_0$ and $\alpha_1$ compared to UEE. Likewise, MAEE also improves the empirical coverage for $\alpha_0$ and $\rho$ under the exponential decay correlation structure, and the results are presented in SFigure 4 and 5.

\section{Analysis of the Washington State EPT Trial}
We apply the cluster-period GEE to analyze cluster-period proportions in the Washington State EPT trial. The focus of this analysis is on estimating the intervention effect and the intraclass correlation structure with respect to the Chlamydia outcomes. We consider the marginal model for cluster-period means $\text{logit}(\mu_{ij})=\beta_j+\delta X_{ij}$, where $\beta_j$ $(j=1,\ldots,5)$ is the period effect and $\exp(\delta)$ is the population-averaged odds ratio. To model the within-cluster correlations, we consider the simple exchangeable structure, as well as the nested exchangeable and exponential decay structures. Of note, the simple exchangeable structure is obtained when we enforce $\alpha_1=\alpha_0$ in the nested exchangeable structure or $\rho=1$ in the exponential decay structure. We do not consider the working independence assumption, because reporting ICCs is considered good practice per the CONSORT extension and useful for planning future SW-CRTs \citep{Hemming2018}.

Table \ref{tb:overall} summarizes the point estimates and bias-corrected standard errors of the marginal mean and correlation parameters from the analysis of all sentinel women. Informed by the simulation study in Section \ref{sec:sim}, we report the BC1 standard error estimates for all marginal mean parameters and the BC2 standard error estimates for all correlation parameters. The estimated odds ratios due to the EPT intervention are $0.868$, $0.867$ and $0.883$, under the simple exchangeable, nested exchangeable and exponential decay correlation models. All 95\% confidence intervals include one. Because the prevalence of Chlamydia is around 6\% and is considered rare, the odds ratio approximates the relative risk. Therefore, interpreting the odds ratio as a relative risk, we conclude that the EPT intervention results in an approximately $12\%$ reduction in Chlamydial infection among women aged between 14 and 25. This finding is consistent with that in \citet{Golden2015} based on generalized linear mixed models. We additionally report the intraclass correlation estimates and their estimated precisions for the Chlamydia outcome on the natural scale of the measurement. The within-period correlation $\hat{\alpha}_0\approx 0.007$, as estimated from both the nested exchangeable and exponential decay model. Under the nested exchangeable model, the between-period correlation $\hat{\alpha}_1\approx 0.004$, which is approximately one half of within-period correlation. The standard error of the between-period correlation is also much smaller than that of the within-period correlation. Under the exponential decay model, the decay parameter is estimated to be $\hat{\rho}\approx 0.7$, suggesting a moderate degree of between-period correlation decay over five periods.

{As an exploratory analysis, we adapt the Correlation Information Criteria \citep[CIC;][]{Hin2009} for the cluster-period GEE and compare the fit of the correlation structures.} Specifically, we define
\begin{equation}\label{eq:cic}
\text{CIC}_{\text{cp}}=\text{trace}\left[\left(\sum_{i=1}^I \bfD_{1i}^T\bfPsi_{1i}^{-1}\bfD_{1i}\right)\bfOmega\bfLambda_{11}
\bfOmega\given[\Big]_{\bftheta=\hat{\bftheta}(\bfR_i),\bfalpha=\hat{\bfalpha}(\bfR_i)}\right],
\end{equation}
where $\bfPsi_{1i}$ is a $J\times J$ working independence covariance, $\bfOmega\bfLambda_{11}\bfOmega$ is the bias-corrected sandwich variance of the marginal mean (BC1), and all parameters are evaluated at the estimates under the assumed correlation structure. According to Theorem \ref{thm:same}, as $I\rightarrow\infty$, the limit of \eqref{eq:cic} is identical to the limit of the usual CIC defined for individual-level GEE in \citet{Hin2009} under the marginal mean model \eqref{eq:mm}, providing some justification for using this metric. {From Table \ref{tb:overall}, while the smallest $\text{CIC}_{\text{cp}}$ corresponds to and favors the exponential decay correlation, the $\text{CIC}_{\text{cp}}$ of the simple exchangeable correlation structure is only larger by a small amount. Future simulation studies are needed to better assess the operating characteristics of $\text{CIC}_{\text{cp}}$ for selecting the optimal correlation structures based on cluster-period GEE analysis of SW-CRTs.}

To explore treatment effect among subgroups, we perform the cluster-period GEE analyses for adolescent girls (aged between 14 and 19) and adult women (aged greater than 19), and present the results in STable 5 and 6. From STable 5, the intervention leads to a more pronounced reduction of Chlamydia infection among adolescent girls compared to the overall analysis. Under the nested exchangeable correlation model, the intervention effect in odds ratio is estimated as $0.780$, and its 95\% CI $(0.625, 0.972)$ excludes one. In other words, the EPT intervention results in $22\%$ reduction in Chlamydia infection among adolescents. Given that adolescents are at high risk for acquiring sexually transmitted diseases and that research on effectiveness of EPT among this population is limited \citep{Gannon-Loew2017}, our subgroup analysis may provide new evidence. For brevity, the comparison between correlation structures among the adolescent population and the analysis of the adult women subgroup are presented in the SM Appendix E.

\section{Discussion}

In the analysis of SW-CRTs with binary outcomes, statistical methods are seldom used to simultaneously obtain point and interval estimates for the intervention effect and the ICCs. An exception is the paired estimating equations approach studied in \citet{LiTurnerPreisser2018}. However, that approach is computationally infeasible with large cluster sizes, which are typically seen in cross-sectional SW-CRTs. To address this limitation, we propose a simple and efficient estimating equations approach based on cluster-period means, which resolves the computational burden of the approach based on individual-level observations. {In practice, one could first attempt an individual-level GEE analysis with an appropriate correlation structure. However, if that procedure becomes computationally intensive, the proposed cluster-period GEE provides a valid workaround. Because standard software could only provide valid intervention effect estimates with cluster-period means in SW-CRTs, we have developed an R package \verb"geeCRT" to implement both the individual-level GEE studied in \citet{LiTurnerPreisser2018} and the cluster-period GEE proposed in this article.}

{Although individual-level analysis has usually been considered more efficient than cluster-level analysis in CRTs,  we have shown in Theorem \ref{thm:same} that the cluster-period GEE and individual-level GEE are asymptotically equally efficient in estimating the treatment effect parameter in cross-sectional SW-CRTs}. The full efficiency of the cluster-period analysis depends on the induced correlation structure, defined in Section \ref{sec:nex} and \ref{sec:ed}. On the other hand, cluster-period analysis of SW-CRTs assuming working independence suffers from statistical inefficiency \citep{Thompson2018}. The numerical study in Section \ref{sec:eff} emphasizes the necessity of carefully characterizing the induced cluster-period correlation when performing a cluster-period analysis. {Finally, the proposed approach enables fast estimation and inference for the correlation parameters, which aligns with the current recommendation in the CONSORT extension to SW-CRTs \citep{Hemming2018}.} The estimating equations method also produces standard errors of the estimated correlations, which can be used to construct interval estimates to further improve planning of future trials.

{Our simulations indicate that the cluster-period GEE can estimate the intervention effect with negligible bias, regardless of bias-corrections to the correlation estimating equations via MAEE. However, MAEE substantially reduces the bias of the ICC estimates. On the other hand, while the bias-corrected sandwich variances can provide nominal coverage for $\delta$ even when $I=12$, inference for ICC parameters appears more challenging (see SM Appendix F for a concise summary of findings). We suggest that $30$ to $40$ clusters may be sufficient for the cluster-period MAEE to provide nominal coverage for $\alpha_0$, which generally agrees with \citet{Preisser2008} using individual-level MAEE. In SW-CRTs, a larger number of clusters may be needed to achieve nominal coverage for the between-period correlation ($\alpha_1$ or $\rho$), which differs from findings in \citet{Preisser2008} for parallel CRTs. This difference highlights the requirement for accurate ICC inference can depend on randomization design (parallel versus stepped wedge). A further reason underlying such a difference is that we have simulated unequal cluster-period sizes, under which the sandwich variance becomes more variable \citep{Kauermann2001}. Fortunately, compared to UEE, the use of cluster-period MAEE can substantially mitigate, if not eliminate, the under-coverage of ICC parameters in small samples.}

{A reviewer has raised the issue of performing cluster-period analysis using GLMMs in cross-sectional SW-CRTs. As explained in SM Appendix G, with binary outcomes, a rigorous cluster-period analysis using GLMMS may proceed with the cluster-period totals, $\sum_{k=1}^{n_{ij}}Y_{ijk}$, which follows a Binomial distribution. The likelihood principle then directly suggests such cluster-period aggregation leads to the same inference of GLMM parameters. However, the interpretation of the treatment effect parameter in GLMMs is conditional on the latent random effects, and therefore applies only to each cluster, or, strictly speaking, to the population with the same value of the unobserved random effects. In contrast, the treatment effect $\delta$ is averaged over all clusters, and has been argued to bear a more straightforward population-averaged interpretation \citep{Preisser2003,LiTurnerPreisser2018}.}

Although the proposed approach is motivated by cross-sectional SW-CRTs, it is equally applicable to parallel cross-sectional longitudinal cluster randomized trials (L-CRTs). In parallel L-CRTs, the intervention effect is parameterized either by the time-adjusted main effect or the treatment-by-time interaction. For both estimands, because cluster-period aggregation implies the same marginal mean model, the proposed GEE approach is valid and can be useful. {Another direction for future research is to extend the proposed approach to analyze closed-cohort SW-CRTs \citep{copas2015designing,LiTurnerPreisser2018,Li2020} and SW-CRTs with continuous recruitment \citep{Grantham2019,hooper2019stepped}. These more recent variants of SW-CRTs have more complex intraclass correlation structures and therefore requires additional considerations in cluster-period analysis.}

{One potential limitation of the current study is that we have only considered a marginal mean model without individual-level covariates. Such an unadjusted mean model originates from \citet{Hussey2007} and has been widely applied for planning and analyzing SW-CRTs; see, for example, the recent review in \citet{LiReview2020}. More often than not, the intraclass correlation structures are also defined with respect to the unadjusted mean models in SW-CRTs \citep{Kasza2017,LiTurnerPreisser2018,Li2020,LiReview2020}. However, covariate adjustment may potentially improve the efficiency in estimating the treatment effect. We plan to carry out future work to integrate individual-level covariates into the cluster-period GEE approach, along the lines of the two-stage framework as in \citet{yasui2004evaluation}.}

\section{Software}
\label{sec5}
An R package for our method, \verb"geeCRT", is available online at CRAN. Sample R code, together with a simulated data example is also available from the corresponding author's GitHub page at \url{https://github.com/lifanfrank/clusterperiod_GEE}.

\section{Supplementary Material}
\label{sec6}

Supplementary material is available online at
\url{http://biostatistics.oxfordjournals.org}.

\section*{Acknowledgments}
Research in this article was partially funded through a Patient-Centered Outcomes Research Institute\textsuperscript{\textregistered} (PCORI\textsuperscript{\textregistered} Award ME-2019C1-16196). The statements presented in this article are solely the responsibility of the authors and do not necessarily represent the views of PCORI\textsuperscript{\textregistered}, its Board of Governors or Methodology Committee. Dr. Preisser has received a stipend for service as a merit reviewer from PCORI\textsuperscript{\textregistered}. Dr. Preisser did not serve on the Merit Review panel that reviewed his project. The authors thank Dr. James P. Hughes for sharing the Washington State EPT study data, and Xueqi Wang for discussions and computational assistance. The authors are also grateful to the Associate Editor and two anonymous reviewers for constructive comments.

\bibliographystyle{biorefs}
\bibliography{refs}

\begin{table}[!p]
\centering
\caption{Mean relative efficiency in estimating the marginal intervention effect $\delta$ when the correlation structure is properly modeled versus when working independence is used in the GEE analyses of cross-sectional SW-CRTs with variable cluster-period sizes bootstrapped from the Washington EPT study. Results are averaged over 1000 bootstrap replicates.}\label{tb:MRE}
\vspace{0.2cm}
\begin{tabular}{cccccccccccc}
\toprule
\multicolumn{6}{c}{True correlation: nested exchangeable} & \multicolumn{6}{c}{True correlation: exponential decay}\\\midrule
\multicolumn{2}{c}{$\alpha_0=0.1$} & \multicolumn{2}{c}{$\alpha_0=0.05$} &
\multicolumn{2}{c}{$\alpha_0=0.02$} & \multicolumn{2}{c}{$\alpha_0=0.1$} & \multicolumn{2}{c}{$\alpha_0=0.05$} &
\multicolumn{2}{c}{$\alpha_0=0.02$}\\
\cmidrule(lr){1-6}\cmidrule(lr){7-12}
$\alpha_1$ & ARE & $\alpha_1$ & ARE & $\alpha_1$ & ARE & $\rho$ & ARE & $\rho$ & ARE & $\rho$ & ARE\\
\midrule
$0.1$ & 44.9 & $0.05$ & 22.5 & $0.02$ & 9.4 & $1$ & 44.9 & $1$ & 22.5 & $1$ & 9.4\\
$0.08$ & 6.5 & $0.04$ & 5.4 & $0.01$ & 2.3 & $0.8$ & 5.7 & $0.8$ & 4.7 & $0.8$ & 3.4\\
$0.06$ & 3.6 & $0.03$ & 3.3 & $0.005$ & 1.8 & $0.5$ & 2.5 & $0.5$ & 2.3 & $0.5$ & 2.0\\
$0.04$ & 2.5 & $0.02$ & 2.4 & $0.002$ & 1.6 & $0.2$ & 1.9 & $0.2$ & 1.8 & $0.2$ & 1.6\\
$0.02$ & 2.0 & $0.01$ & 1.9 & $0.001$ & 1.6 & $0.05$ & 1.9 & $0.05$ & 1.8 & $0.05$ & 1.6\\
\bottomrule
\end{tabular}
\end{table}

\begin{table}
\centering
\caption{Percent relative bias of model parameters as a function of number of clusters $I$ under two different correlation structures: nested exchangeable (NE) and exponential decay (ED). The cluster-period sizes are randomly drawn from DiscreteUniform$(50,150)$ and the results are based on $3000$ simulations.}\label{tb:bias}
\vspace{0.2cm}
\begin{tabular}{llrrrrrr}
\toprule
& & \multicolumn{2}{c}{Bias $\hat{\delta}$} & \multicolumn{2}{c}{Bias $\hat{\alpha}_0$} &
\multicolumn{2}{c}{Bias $\hat{\alpha}_1$ or $\hat{\rho}$}\\
\cmidrule(lr){3-4}\cmidrule(lr){5-6}\cmidrule(lr){7-8}
& & UEE & MAEE & UEE & MAEE & UEE & MAEE\\
\cmidrule{3-8}
\multicolumn{2}{c}{Correlation structure} & \multicolumn{6}{c}{$I=12$}\\
\cmidrule{1-8}
\multirow{2}{*}{NE$(\alpha_0,\alpha_1)$}
& $(0.03, 0.015)$ & 0.5 & 0.5 & -13.9 & -0.5 & -10.7 & -1.5 \\
& $(0.1, 0.05)$ & 2.2 & 2.2 & -9.9 & 0.9 & -8.7 & 0.3 \\
\multirow{2}{*}{ED$(\alpha_0,\rho)$}
& $(0.03, 0.8)$ & 0.2 & 0.2 & -12.6 & -0.1 & -1.8 & -4.0 \\
& $(0.1, 0.5)$ & 2.8 & 2.8 & -9.5 & 1.6 & -2.1 & -3.0 \\
\cmidrule{1-8}
\multicolumn{2}{c}{Correlation structure} &  \multicolumn{6}{c}{$I=24$}\\
\cmidrule{1-8}
\multirow{2}{*}{NE$(\alpha_0,\alpha_1)$}
& $(0.03, 0.015)$ & 0.3 & 0.3 & -6.8 & -0.2 & -4.6 & 0.0 \\
& $(0.1, 0.05)$ & 0.3 & 0.3 & -5.2 & 0.1 & -4.4 & 0.1 \\
\multirow{2}{*}{ED$(\alpha_0,\rho)$}
& $(0.03, 0.8)$ & 0.2 & 0.2 & -6.4 & -0.2 & -0.4 & -1.4 \\
& $(0.1, 0.5)$ & 0.6 & 0.6 & -5.3 & 0.1 & -1.5 & -1.9 \\
\cmidrule{1-8}
\multicolumn{2}{c}{Correlation structure} &  \multicolumn{6}{c}{$I=36$}\\
\cmidrule{1-8}
\multirow{2}{*}{NE$(\alpha_0,\alpha_1)$}
& $(0.03, 0.015)$ & 0.5 & 0.5 & -5.0 & -0.6 & -4.2 & -1.2 \\
& $(0.1, 0.05)$  & 1.1 & 1.1 & -3.7 & -0.2 & -3.8 & -0.8 \\
\multirow{2}{*}{ED$(\alpha_0,\rho)$}
& $(0.03, 0.8)$ & 0.4 & 0.4 & -4.9 & -0.8 & -0.7 & -1.3 \\
& $(0.1, 0.5)$  & 1.3 & 1.3 & -3.6 & 0.0 & -1.4 & -1.7 \\
\bottomrule
\end{tabular}
\end{table}

\begin{table}
\centering
\caption{Parameter estimates of marginal mean and correlation parameters from the overall analysis of Washington State EPT Trial using MAEE. Standard error of the marginal mean parameters are based on BC1 and standard error of the intraclass correlation parameters are based on BC2. All standard error estimates are reported in the parenthesis.}\label{tb:overall}
\vspace{0.2cm}
\begin{tabular}{lccc}
\toprule
& Simple exchangeable & Nested exchangeable & Exponential decay \\
\cmidrule{2-4}
\multicolumn{4}{l}{\emph{Marginal mean}}\\
$\beta_1$ (period 1) & -2.443 (0.091) & -2.446 (0.095) & -2.437 (0.095) \\
$\beta_2$ (period 2) & -2.454 (0.091) & -2.439 (0.083) & -2.444 (0.089) \\
$\beta_3$ (period 3) & -2.535 (0.094) & -2.495 (0.100) & -2.508 (0.100) \\
$\beta_4$ (period 4) & -2.609 (0.106) & -2.606 (0.115) & -2.613 (0.115) \\
$\beta_5$ (period 5) & -2.537 (0.145) & -2.535 (0.128) & -2.552 (0.131) \\
$\delta$ (treatment) & -0.141 (0.092) & -0.142 (0.090) & -0.124 (0.087) \\
\cmidrule{2-4}
\multicolumn{4}{l}{\emph{Intraclass correlation}}\\
$\alpha_0$ & .0051 (.0016) & .0072 (.0039) & .0070 (.0039)\\
$\alpha_1$ & -- & .0038 (.0015) & -- \\
$\rho$ & -- & -- & .7157 (.2962)\\
\cmidrule{2-4}
\multicolumn{4}{l}{\emph{Correlation selection criteria}}\\
$\text{CIC}_{\text{cp}}$ & 16.12 & 16.53 & 16.10\\
\bottomrule
\end{tabular}
\end{table}

\end{document}